\documentstyle[epsf]{article}
\def\nl{\hfil\break}

\def\npb{{ \sl Nucl. Phys. }}
\def\lesssim{\mathrel{\raise.3ex\hbox{$<$\kern-.75em\lower1ex\hbox{$\sim$}}}}
\def\half{{1\over2}}
\def\bfr#1{{#1}_{\bf r}}
\def\tonez{T_1^0}
\def\ttwoz{T_2^0}
\begin{document}
\centerline{\Large Classical $\phi^4$ Lattice Field Theory in 
   Strong Thermal Gradients}\nl
\centerline{  Kenichiro Aoki$^a$ \ and Dimitri Kusnezov$^b$ }
\centerline{$^a$\it Dept. of
  Physics, Keio University, {\it 4---1---1} Hiyoshi, Kouhoku--ku,
  Yokohama 223--8521, Japan}
\centerline{$^b$\it Center for Theoretical Physics,
  Sloane Physics Lab, Yale University, New Haven, CT\ 06520-8120}
\small
\centerline{\bf Abstract}
\begin{quotation}
  The dynamics of  classical $\phi^4$ theory under weak and strong
  thermal gradients is studied.  We obtain the thermal conductivity of
  the theory including its temperature dependence.  Under moderately
  strong thermal gradients, the temperature profiles become visibly
  non-linear, yet the phenomenon can be understood using the linear
  response theory.  When we move further away from equilibrium, we
  find that the linear response theory eventually breaks down, and the
  concept of local equilibrium also fails.
\end{quotation}
\section{Introduction}
\label{sec:intro}
It is widely accepted that the non-equilibrium dynamics of field
theories has many important but difficult issues yet to be understood.
The motivation for considering such theories hardly needs to be
stressed --- non-equilibrium situations are ubiquitous --- from
processes in inflation or baryogenesis in the early universe, 
transport processes in condensed matter, to the  possible
states of hadronic matter in heavy ion collisions, such as
quark--gluon plasma, disoriented chiral condensates and color
superconducting states.  These phenomena all involve the
non-equilibrium dynamics in some essential manner.

The particular non-equilibrium problem we study is the behavior of a
field theory when various temperature boundary conditions are imposed
on the boundaries of the theory.  The physical properties such as the
temperature profile, $T(x)$, pressure or entropy inside the boundaries,
are determined {\it dynamically}.  It is {\it \`a priori} not
clear whether the system thermalizes when strong thermal gradients are
present, and we would like to clarify the situation.  It would be
preferable to compute the physical properties of the theory within an
analytic field theory, yet such an approach seems difficult:  One can
compute the transport coefficients within  linear response theory,
since such a computation is performed in {\it equilibrium}, yet even
then, its region of applicability is unclear and in principle could even be
null, as was found in some cases\cite{div-transport}.  Within the
linear regime, we might try to approach the problem directly using
thermofield dynamics, for instance, but it seems difficult to do so
without imposing some assumptions on the dynamics of the theory.
Beyond the linear regime, it seems fair to say that the problem is
very difficult.

In this work, we impose various temperature boundary conditions
on massless $\phi^4$ theory in (1+1) and (3+1) dimensions and
study the behavior of the theory in the steady state.  We make
use of numerical methods to compute physical quantities of
interest.  We shall be interested in such questions as the
validity of the linear response theory and the region of its
applicability, or the possibility of the failure to achieve
thermalization.  Let us point out the limitations of our current
approach: The theory we work with is classical and it is
formulated on a lattice.  However, apart from this, we make no
assumptions on the dynamics of the theory and we compute
physical observables from first principles.  Also, it should be
clear from our approach that our methods are applicable to other
field theories as well, except perhaps for the need for more
computational time in more complicated problems.  We choose the
$\phi^4$ theory since it is a prototypical field theory and it
appears in various contexts in many areas of physics.  It should
also be pointed out that classical approximations to quantum
field theories have been studied for some time and while far
from trivial, a basic understanding of the relation of the
classical theory to the quantum one for high temperatures does
exist \cite{classical-approx}.  In addition, the classical
theory is of interest on its own right and we believe that the
understating of its dynamics is essential, if not necessary, to
the understanding of the quantum theory.
\section{The model}
\label{sec:model}
The Lagrangian of the model we study, the massless $\phi^4$ theory, is
in the continuum,
\begin{equation}
  \label{lag}
  -\c L=\left(\partial_\mu\tilde\phi\right)^2
  +{{\tilde g}^2\over4}\tilde\phi^4
\end{equation}
We may scale out the dimensionful variables and the coupling using the
rescalings ${\bf r}=\tilde {\bf r}/a, t=\tilde t/a, \phi_{\bf r}=a\tilde
g\tilde \phi(\tilde {\bf r},\tilde t)$ with $a$ being the lattice
spacing.  We obtain the Hamiltonian on the lattice, $H$,
\begin{equation}
  \label{ham}
  H=\sum_{\bf r}\left[\half\bfr\pi^2+\half
    (\vec\nabla\phi)_{\bf r}^2+{1\over4}\phi_{\bf r}^4\right],\qquad
  0\leq x\leq  L, 0\leq y,z\leq L'
\end{equation}
Here, $(\nabla_k\phi\bfr)=\phi_{{\bf r}+{\bf e}_k}-\bfr\phi$ where
${\bf e}_k$ is the unit vector in the $k$-th direction.

We  thermostat the boundaries at temperatures $T(x=0)=T_1^0$ and
$T(x=L)=T_2^0$.  In (3+1) dimensions, we impose  periodic boundary
conditions in the $y,z$ directions.  In this manner, statistical
averages of observables in equilibrium or non-equilibrium steady states
are equivalent to time averages in the long-time limit. The dynamics of
the system is purely that of the $\phi^4$ theory inside the boundaries
$0<x<L$.  
The thermostats are provided using
the ``global demons'' of \cite{thermostats}, a non-Hamiltonian
generalization of the Nos\'e--Hoover approach\cite{nose-hoover}.

When the boundary temperatures are equal, $\tonez=\ttwoz$, we recover
the equilibrium ensemble, as we should.  The temperature inside the
system is the same as the boundary temperature, $T({\bf r})=\tonez\ 
(0<x<L)$ and the distribution of the momentum $\pi_{\bf r}$ at any
site is Maxwellian.  As we make the boundary temperatures different,
we find the linear regime and then the non-linear regime.  We now
discuss the behavior of the system under these conditions.
\section{Weak Thermal Gradients: $\tonez\lesssim\ttwoz$}
\label{sec:linear}
When the boundary temperature difference is small, a linear
temperature profile emerges.  An example of such a profile is shown in
Fig. 1.
We find that the distributions of the momenta are thermal at any of
the points inside and outside the boundaries.  Thermalization of the
region outside the boundaries is established through the direct
coupling to the thermostats while the thermalization of the sites
inside the boundaries is established dynamically.
\begin{figure}[htbp]
  \begin{center}
    \leavevmode
    \epsfysize=4.5cm\epsfbox{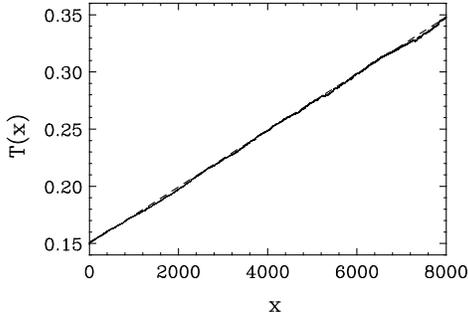}
    \hskip1cm\parbox[b][4.5cm][c]{6cm}{
    \caption{\small
      A linear temperature profile for $L=8000,
      \tonez=0.15,\ttwoz=0.35$, compared to the linear fit (dashes). }     }
    \label{fig:linear}
  \end{center}
\end{figure}

\begin{figure}[htbp]
  \begin{center}
  \leavevmode
  \epsfysize=5cm\epsfbox{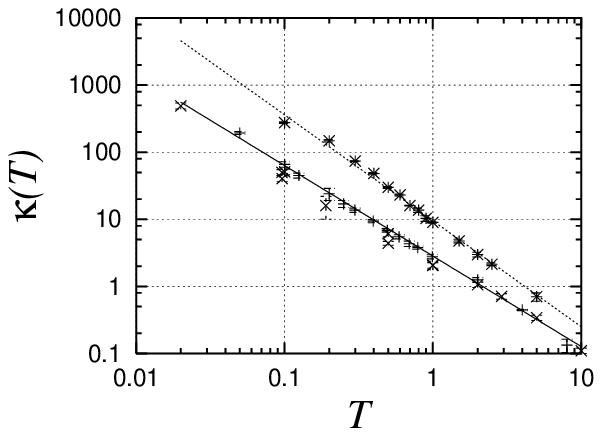}\hskip1.5cm
  \epsfysize=5cm\epsfbox{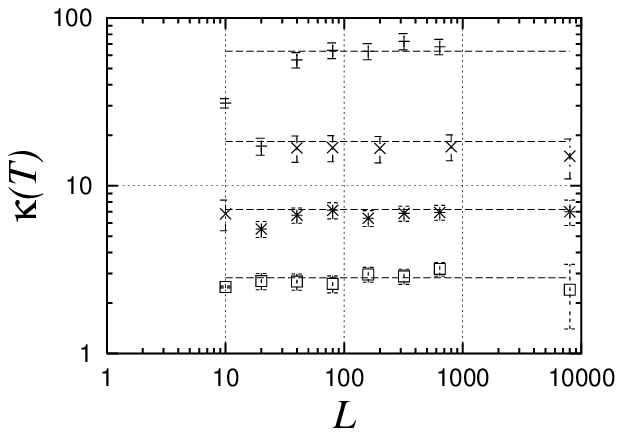}
    \caption{\small 
      The temperature dependence of the thermal conductivity (left):
      The direct measurement in (1+1)-d ($\times$) and (3+1)-d ($*$)
      obtained through the application of Fourier's law and the
      Green--Kubo predictions in (1+1)-d ($+$) are plotted.  These two
      independent methods clearly lead to the same result.  The volume
      dependence of the thermal conductivity in (1+1)-d for
      the temperatures 0.1 ($+$), 0.25 ($\times$), 0.5 ($*$) and 0.1
      ($\Box$).   The dashed lines denote the values of thermal
      conductivity predicted by Eq.~(\ref{tc}).  We see that the bulk
      limit is reached for reasonably small lattices in this
      temperature range.}
  \label{fig:tc-t}
\end{center}
\end{figure}

Using these types of linear profiles, we have obtained the
thermal conductivity of the system for various temperatures and
various lattice sizes.  We expect that the linear region should
be well described by Fourier's law,
\begin{equation}
  \label{fourier}
  \langle{\c T}^{01}\rangle_{\scriptstyle NE}=-\kappa(T)\nabla T,\qquad
  {\rm where}\quad 
  \c T^{01}_k=-\pi_k(\nabla \phi)_k ,
\end{equation}
$\langle \cdots\rangle_{\scriptstyle NE}$ is the non-equilibrium
average, and
$\c T^{01}$ is the heat flux in our theory.  While it is quite
possible to extract a value for the thermal conductivity from systems
like the one shown in Fig.~1, we have extracted the thermal
conductivity for a given temperature from systems obtained by varying
the temperature difference around the given temperature.  We find that
the thermal conductivity has a well defined bulk limit.  In other
words, it remains constant when the lattice size is increased for
moderately large lattices.  The thermal conductivity we find is
described by a power law as shown on Fig.~\ref{fig:tc-t}, both in
(1+1) ($\times$) and (3+1) ($*$)
dimensions:
\begin{equation}
  \label{tc}
  \kappa(T) = {A\over T^\gamma},\qquad
  \cases{\gamma=1.35(2),\ A=2.83(4) & (1+1) dimensions\cr
    \gamma=1.58(4),\ A=9.5(5) & (3+1) dimensions\cr}
\end{equation}

The thermal conductivity may also be computed in a completely
different manner in {\it equilibrium}, by using the Green--Kubo
formula for the thermal conductivity,
\begin{equation}
  \label{green-kubo}
  \kappa(T)={1\over T^2}\int dx\,dt\,
  \left\langle\c T^{01}(x_0,t_0)T^{01}(x,t)\right\rangle_{eq} .
\end{equation}
Applying the formula to our lattice theory, we obtain the
thermal conductivity for various temperatures which agree well
with our direct measurements, as we can see in
Fig.~\ref{fig:tc-t} (left, $+$).  While this might seem obvious,
it should be pointed out that the integrands in the Green--Kubo
formula have ``long time tails'' and the integrand was found to
be divergent in various low dimensional systems
\cite{div-transport}.  In our theory, the long time tails do
exist in (1+1) dimensions up to $\sim10\tau$, where $\tau$ is
the mean free time.  These transient tails do have the expected
behavior of $t^{-1/2}$, which, if it continued, would lead to a
divergent integral in the Green--Kubo formula.  In our case, the
long time tails are transient and the integrand decays much
faster for $t\gg10\tau$ leading to a finite transport
coefficient.
\section{Strong Thermal Gradients: $\tonez\ll\ttwoz$}
\label{sec:non-linear}
\begin{figure}[htbp]
   \begin{center}
    \leavevmode
    \epsfysize=5cm\epsfbox{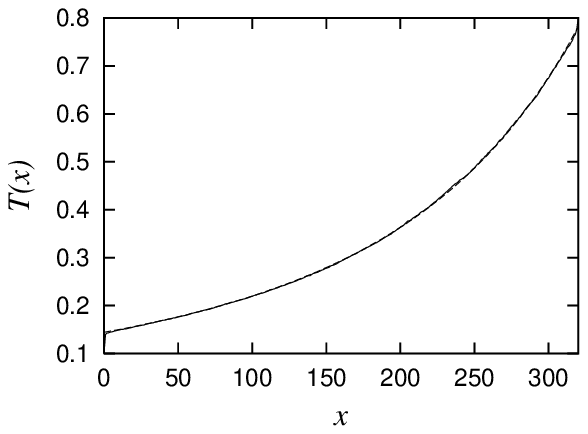}
    \hskip1cm
    \epsfysize=5cm\epsfbox{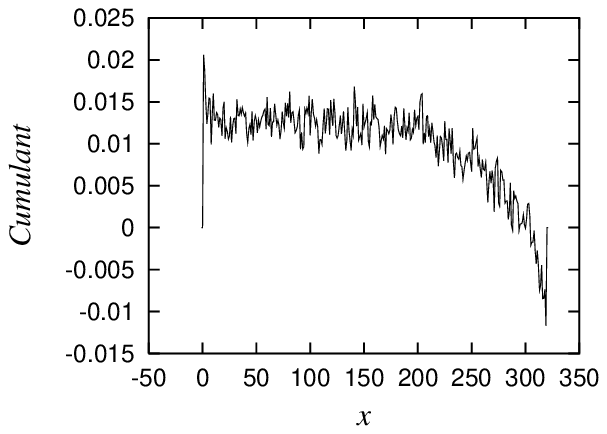}
    \caption{\small
      Left figure: A temperature profile with visible curvature
      (solid) with its fit (dashed).  
      In this example, $L=320, (T^0_1,T^0_2)=(0.1,0.8)$.  We
      also see that temperature jumps develop at the boundaries.
      Right figure: The cumulant
      $\langle\pi_k^4\rangle/(3\langle\pi_k^2\rangle^2) - 1$ for
      the system.  We see that the deviations from the
      equilibrium value is of the order of a percent or so and
      the concept of local equilibrium applies sufficiently
      well. }
    \label{fig:curved-profile}
   \end{center}
\end{figure}

In the regime where the two boundary temperatures are substantially
different, the thermal profiles become visibly curved.  An example of
such a profile is show in Fig.~\ref{fig:curved-profile} as the solid
curve (dashed curved will be explained later).  Also, another feature
that emerges is that jumps in the temperature arises at the
boundaries; namely, the temperature obtained by extrapolating the
temperature profile inside does not match the boundary thermostat
temperatures.  We would like to understand the physics behind these
features.

The boundary temperature jumps can be understood quantitatively using
kinetic theory ideas.  We dispense with the details here which can be
found in \cite{ak1}.  We should point out, however, that these jumps
are physical and are observed generically in real
systems\cite{jump-exp}\ as well as simulations\cite{jump-sim} for
systems that are far away enough from the equilibrium.  In any case,
it should be emphasized that as long as the boundary thermostats are
thermalizing the degrees of freedom outside the boundaries --- which
we indeed do confirm --- what happens within the boundaries is
determined {\it dynamically} only by the degrees of freedom of the
$\phi^4$ theory.

Let us now move on to the curved temperature profiles: We first note
that even within the context of linear response theory, the
temperature profile will become curved since the thermal conductivity
is a non--trivial function of the temperature, as in Eq.~(\ref{tc}).
If we assume that this is the only cause of the non--linearity of the
profile, we obtain
\begin{equation}
  \label{eq:profile}
  T(x)=T_1\left[1-{x\over L}+ \left(T_2\over T_1
    \right)^{1-\gamma}
      {x\over L}\right]^{1\over1-\gamma}  
\end{equation}
where $T_{1,2}$ are the extrapolated boundary temperatures which in
general can be different from the thermostat temperatures,
$T^0_{1,2}$, due to the existence of the boundary temperature jumps.
We find that as long as the temperature gradient is not too large, the
temperature profile is well explained by this formula.  An example of
such a fit is shown in Fig.~\ref{fig:curved-profile} (left) as a
dashed curve --- in fact, the fit is barely distinguishable from the
observed thermal profile except at the boundaries.  It should be noted
that $\gamma$ in Eq.~(\ref{eq:profile}) is determined independently
from systems near and at equilibrium.  We have
also looked at the momentum distributions in these systems at various
sites and find that they are quite close to their equilibrium gaussian
distributions.  To analyze this quantitatively, one may look at the
higher order cumulants of $\pi_k$,
\begin{equation}
  \label{cumulant}
  {\langle\pi_k^4\rangle\over3\langle\pi_k^2\rangle^2}  -1,
  {\langle\pi_k^6\rangle\over15\langle\pi_k^2\rangle^3}  -1,
  \ldots  ,\qquad k=0,1,\ldots L
\end{equation}
The spatial dependence of the fourth order cumulant is shown in
Fig.~\ref{fig:curved-profile}, where we see that the deviation from
the equilibrium value, $0$, is at a percent level.  So in conclusion,
there exists a regime where the thermal profiles are curved, yet both
the concept of local equilibrium and linear response theory apply
quite well.  \`A priori, this needed {\it not} be the case.

Let us briefly discuss what happens when we make the temperature
difference larger and larger.  We find that linear response predictions
break down and that the profiles are no longer described by the
equation Eq.~(\ref{eq:profile}).  It is tempting to interpret this as
a ``non-linear'' response of the system involving higher order
derivatives of the temperature, such as $\nabla ^3 T,\ \nabla
T\nabla^2 T$,... .  However, before we apply these ideas, we should check
that local equilibrium is achieved and that the usual concept of
temperature applies.  In our approach, whether local equilibrium is
achieved or not is a dynamical question that can be answered
unambiguously.  We find that when the linear response does not apply,
neither does local equilibrium hold.  So, in fact, the situation
is much more complicated; while non-linear response might indeed
exist, at least for the theory at hand, the failure of local
equilibrium needs to be taken into account simultaneously.
\section{Discussions}
\label{sec:disc}
The dynamics of the classical lattice $\phi^4$ theory was studied
under weak and strong thermal gradients from first principles.  In
equilibrium, the Green--Kubo formula was applied to derive the thermal
conductivity.  This was found to agree with that obtained using the
Fourier's law for the system under weak thermal gradients.  
To our knowledge, this is the first time the non--trivial temperature
dependence of the thermal conductivity computed from first
principles over a number of decades for any system.  
For moderately strong gradients, the linear response theory was found
to be quite applicable even though the thermal profiles were visibly
curved.  For even stronger gradients, linear response theory ceases to
hold and local equilibrium breaks down also at the same time.  While
we did not have time to discuss this here, we have also studied other
quantities in the theory, such as the heat capacity, entropy and the
speed of sound and have elucidated how they are related to each other,
leading to a comprehensive understanding of the dynamics \cite{ak2}.

Clearly, much remains to be done: It would be interesting to study the
dynamics of the theory which exhibits spontaneous symmetry breaking.
In this case, the phase boundary will emerge dynamically and this can
be analyzed within the current approach.  We would also like to
understand the dynamics of other theories, such as Yang--Mills
theories under thermal gradients.  Since our methods are not
restricted to the steady state case, another problem of import is that
of transient phenomena, such as thermalization.  An important point
which needs to be investigated is the relation of the physical
quantities in the lattice theory to those in the continuum theory.  In
a related question, we would like to understand the quantum effects
and understand how the classical theory can be ``matched'' to the
classical theory.  Work is in progress in these areas.

\end{document}